# THE INTERPLAY OF CURVATURE AND VORTICES IN FLOW ON CURVED SURFACES

S. REUTHER[†] AND A. VOIGT[‡]

**Abstract.** Incompressible fluids on curved surfaces are considered with respect to the interplay between topology, geometry and fluid properties using a surface vorticity-stream function formulation, which is solved using parametric finite elements. Motivated by designed examples for superfluids, we consider the influence of a geometric potential on vortices for fluids with finite viscosity and show numerical examples in which a change in the geometry is used to manipulate the flow field.

**1. Introduction.** We consider two-dimensional (2D) films, e.g. thin films whose thickness is much smaller than their lateral extension. Such systems can often easily bend and we are interested in the question how an imposed geometric deformation influences the internal structure of the film. Such an interplay between geometry and internal structure is well studied for condensed matter systems, see [3]. The most illustrative example is probably the structure of a soccer ball, with 12 pentagons, serving as topological defects. Here, the favored regular order of hexagons, which easily tile a flat surface, cannot be extended throughout the surface of the sphere. In technologically more relevant examples of soft materials, such defects are the key for chemical functionalization and provide the opportunity for the design of novel materials [15].

Less explored is an analogy for fluid films. Here, the defects are vortices, which interact with the geometry. In [23] this is analyzed for superfluidic films, fluids with zero viscosity and zero entropy. Thin layers of liquid helium are used as a model system, which can be described by a simple scalar field theory for the superfluidic wave function on the surface. We will here concentrate on the more subtle case of fluids with nonzero viscosity and consider a numerical approach for an incompressible surface Navier-Stokes equation for which we analyze the interplay between the geometry and the vortices of the flow field.

The paper is organized as follows: In §2 we will introduce basic concepts to describe the interplay of geometry and defects. Section §3 is devoted to the incompressible surface Navier-Stokes equation. In §4 we extend our surface vorticity-stream function formulation [17] to evolving surfaces, introduce the numerical approach and consider various examples. Conclusions and an outlook towards biological structures and technological applications are provided in §5.

**2. Basic concepts – geometric potential.** One mechanism for the appearance of defects is a topological constraint. If we triangulate the considered surface we can use the Gauss-Bonnet theorem to establish a relation between the triangulation and the Gaussian curvature $K$

$$V + F - E = \frac{1}{2\pi} \int_\Gamma K \, d\Gamma \tag{2.1}$$

[†]Department of Mathematics, TU Dresden, 01062 Dresden, Germany (sebastian.reuther@tu-dresden.de)
[‡]Department of Mathematics and Center of Advanced Modeling and Simulation, TU Dresden, 01062 Dresden, Germany (axel.voigt@tu-dresden.de)

with the number of vertices $V$, the number of faces $F$, the number of edges $E$ of the triangulation and the closed surface $\Gamma$. Although the Gaussian curvature is a local geometric property, when integrated over $\Gamma$ it becomes a topological invariant, the Euler characteristic

$$\frac{1}{2\pi} \int_\Gamma K \, d\Gamma = \xi(\Gamma). \tag{2.2}$$

For a sphere with radius $R$, the surface area is $4\pi R^2$ and the Gaussian curvature is $1/R^2$, which gives $\xi(S^2) = 2$. For each vertex $i$ of the triangulation, we can assign a coordination number $C_i$, counting the number of edges at that point. Each edge connects two vertices. On the other hand, $i$ is also a vertex of $C_i$ faces, where each connects three vertices. The contribution of vertex $i$ thus can be written as $1 - C_i/2 + C_i/3 = (6 - C_i)/6$. Thus, vertices with a coordination number $C_i = 6$ do not contribute and we can sum over all defects ($C_i \neq 6$) and obtain

$$\sum_i \frac{\text{ind}_S(\mathbf{d}_i)}{6} = \xi(\Gamma) \tag{2.3}$$

with $\text{ind}_S(\mathbf{d}_i) = 6 - C_i$ the defect charge at defect position $\mathbf{d}_i$. For our soccer ball example from the introduction, with each face replaced by a vertex, we thus find 12 points with charge $+1$.

Similar arguments hold for a vector field on a closed surface. Also in this case the Euler characteristic can be used to understand the defects. Here, the Poincaré-Hopf theorem shows, that any continuous vector field on a sphere must have at least two $+1$ defects or one $+2$ defect. Consider e.g. the lines of latitude on the globe that naturally create two vortices at the north and south pole. Here, the charge of a defect is no longer determined by it's coordination number, but by it's winding number, which is the algebraic sum of the number of revolution of the vector field along a small counterclockwise oriented curve around the defect. We obtain

$$\sum_i \text{ind}_V(\mathbf{d}_i) = \xi(\Gamma) \tag{2.4}$$

with index/winding number $\text{ind}_V(\mathbf{d}_i)$. For the example on the globe we have two $+1$ defects. For a more detailed discussion we refer to e.g. [13].

The topology of the surface is one source for the appearance of defects and the total topological charge of all defects is a conserved quantity. However, the defect positions are not determined by topology but result from defect-defect interactions and other sources, such as geometry and dynamics. On a more general surface with a varying Gaussian curvature, each defect experiences an additional geometric potential, which reflects the broken translational invariance of the surface and the type of order in the film or the alignment of the vector field with the surface. In all these cases, defects can be described by an effective free energy interacting among each other [25]. This geometric interaction is given by

$$\mathcal{E}(\mathbf{d}_i) = -2\pi\kappa \, \text{ind}(\mathbf{d}_i) \left(1 - \frac{\text{ind}(\mathbf{d}_i)}{2}\right) U_G(\mathbf{d}_i) \tag{2.5}$$

with elastic stiffness $\kappa$ and geometric potential $U_G$ determined by the surface Poisson equation

$$\Delta_\Gamma U_G = K, \tag{2.6}$$

with Laplace-Beltrami operator $\Delta_\Gamma$. The linear term in eq. (2.5) arises from the geometric frustration of a vector field, while the quadratic term originates from the distortion of a vortex's own flow pattern by the geometry. This self-interaction is analyzed in detail for superfluid helium in [23]. For this case, analytic expressions can be derived for the interaction of vortices, showing e.g. a repulsion of vortices from positive curvature and an attraction to negative curvature regions.

While this mesoscopic approach to describe the system in terms of the interaction of defects is very efficient and successful in determining stable configurations, it neglects other influences for the appearance of defects, which are caused by dynamics. We therefore consider a full dynamic model and use the described approach only to justify the chosen examples and to demonstrate the interaction of the flow field with the geometry. Instead of a superfluid, where dissipation mechanisms of a conventional fluid are absent, we consider an incompressible surface Navier-Stokes equation and show that the same interactions can also be found for that case.

**3. Incompressible surface Navier-Stokes equation.**

**3.1. Model derivation.** Using a generalized Laplacian of a vector field on a surface, the incompressible surface Navier-Stokes equation reads

$$\partial_t \mathbf{v} + \mathbf{v} \cdot \nabla_\Gamma \mathbf{v} = -\nabla_\Gamma p + \mu(\Delta_\Gamma^B \mathbf{v} + K\mathbf{v}) \tag{3.1}$$

$$\nabla_\Gamma \cdot \mathbf{v} = 0 \tag{3.2}$$

with velocity $\mathbf{v}$ defined in the tangential space with components $(v_1, v_2)$ corresponding to the unit vectors $\mathbf{e}_1(\mathbf{x})$ and $\mathbf{e}_2(\mathbf{x})$, which are defined to be perpendicular to the surface normal $\mathbf{n}(\mathbf{x})$ and each other for each $\mathbf{x} \in \Gamma$, pressure $p$ and surface viscosity $\mu$. $\Delta_\Gamma^B$ is the Bochner or rough Laplacian. Besides this generalization of the Laplacian also the Gaussian curvature $K$ enters, which is the reason for the geometric potential discussed above. This formulation, but without the Gaussian curvature term, has been used in [22] to formulate the Navier-Stokes equation on a manifold. An alternative formulation is

$$\partial_t \mathbf{v} + \mathbf{v} \cdot \nabla_\Gamma \mathbf{v} = -\nabla_\Gamma p + \mu(\Delta_\Gamma^R \mathbf{v} + 2K\mathbf{v}) \tag{3.3}$$

$$\nabla_\Gamma \cdot \mathbf{v} = 0 \tag{3.4}$$

with $\Delta_\Gamma^R$ the Laplace-de Rham operator or Hodge-de Rham Laplacian. This formulation is used in [4], but again without the Gaussian curvature term. The correct formulation has been considered in the mathematical literature in [8, 14]. The equation is also related to the Boussinesq-Scriven constitutive law for the surface viscosity in two-phase flow problems [20, 21, 2] and biomembrane problems [12, 1, 9]. However, numerical approaches are restricted until recently to simplified (e.g. spherical or axis-symmetric) geometries.

In [17] the surface vorticity-stream function formulation is introduced for the incompressible surface Navier-Stokes equation, which follows as in 2D by considering the velocity $\mathbf{v}_3 = (v_1, v_2, 0)$ in the coordinate system $(\mathbf{e}_1(\mathbf{x}), \mathbf{e}_2(\mathbf{x}), \mathbf{n}(\mathbf{x}))$ and $\mathbf{v}_3 = \nabla \times \psi$ with the surface stream function $\psi$. We end up with a scalar surface partial differential equation

$$\partial_t \Delta_\Gamma \psi + J(\psi, \Delta_\Gamma \psi) = \mu(\Delta_\Gamma^2 \psi + 2\nabla_\Gamma \cdot (K\nabla_\Gamma \psi)) \tag{3.5}$$

where $J(\psi, \Delta_\Gamma \psi) = -(\nabla \times \psi) \cdot \nabla_\Gamma \Delta_\Gamma \psi$ is the Jacobian. To numerically solve the equation, we rewrite eq. (3.5) as a system of two second order equations

$$\partial_t \phi + J(\psi, \phi) = \mu(\Delta_\Gamma \phi + 2\nabla_\Gamma \cdot (K\nabla_\Gamma \psi)) \tag{3.6}$$
$$\phi = \Delta_\Gamma \psi \tag{3.7}$$

with surface vorticity $\phi$.

**3.2. Numerical approach.** A parametric finite element approach is considered to solve the problem. We here follow the general approach, described in [5, 6, 24], which is already tested for the considered problem in [17]: Let $\Pi_h$ be a surface triangulation of $\Gamma$ of mesh size $h$ such that

$$\Gamma_h = \bigcup_{Z \in \Pi_h} Z$$

is an interpolation of $\Gamma$ and let $\mathbb{T}_\tau$ be a uniform partition of the time interval $(0, T]$ of mesh size $\tau$. We define the discrete time derivative $d_\tau v^m := (v^m - v^{m-1})/\tau$ and introduce the surface finite element space

$$V_h = \{v_h \in H^1(\Gamma_h) \mid v_{h|Z} \in P^1(Z), \; \forall Z \in \Pi_h\}.$$

Thus, the surface finite element approximation reads: Find $(\phi^m, \psi^m) \in V_h \times V_h$ such that for all $(\alpha, \beta) \in V_h \times V_h$

$$(d_\tau \phi^m, \alpha) + (J(\psi^{m-1}, \phi^m), \alpha) = \mu(-\nabla_\Gamma \phi^m - 2K\nabla_\Gamma \psi^m, \nabla_\Gamma \alpha)$$
$$(\phi^m, \beta) = -(\nabla_\Gamma \psi^m, \nabla_\Gamma \beta).$$

Within this semi-implicit discretization, we assume $K$ to be given analytically or computable at the required accuracy, see Appendix A.

**3.3. Simulation results.** We first come back to the topological defects on a sphere and study their interaction. We consider two $+1$ defects (vortices), with initial condition, such that they are not at opposite positions. The vortices are known to repel each other and their interaction energy depends linearly on the vortex separation distance [18]. Fig. 1 shows the dynamics for a viscosity $\mu = 1.0$ towards a state, where they are maximally separated.

Here, the vortices approach the maximal separation directly. However, this is no longer the case for a reduced viscosity. Fig. 2 shows the vortex trajectories for $\mu = 0.01$ as well as the dependency of the time needed to reach the stationary state, in which the vortices have a separation distance of $\pi$, on the viscosity. For less viscous fluids, the vortices spiral towards the stationary state and as larger the viscosity as faster the stationary state is reached.

We now consider a geometry which is topologically equivalent to the sphere, but with varying Gaussian curvature. We use a common test case in computer graphics, the Stanford bunny. To make flow simulations on this geometry feasible, the original mesh had to be improved and the surface had to be smoothed to remove sharp corners. The obtained geometry still contains regions with large positive and negative Gaussian curvature. We start our simulation with noise as initial condition and let the flow field evolve. Fig. 3 shows the reached stationary state, with high velocity differences and three $+1$ defects (vortices) and one $-1$ defect (saddle), see Fig. 4 for details. We thus obtain $\sum_i \mathrm{ind}_V(\mathbf{d}_i) = 1 + 1 + 1 - 1 = 2 = \xi(\Gamma)$ and therefore a different realization of the Poincaré-Hopf theorem.

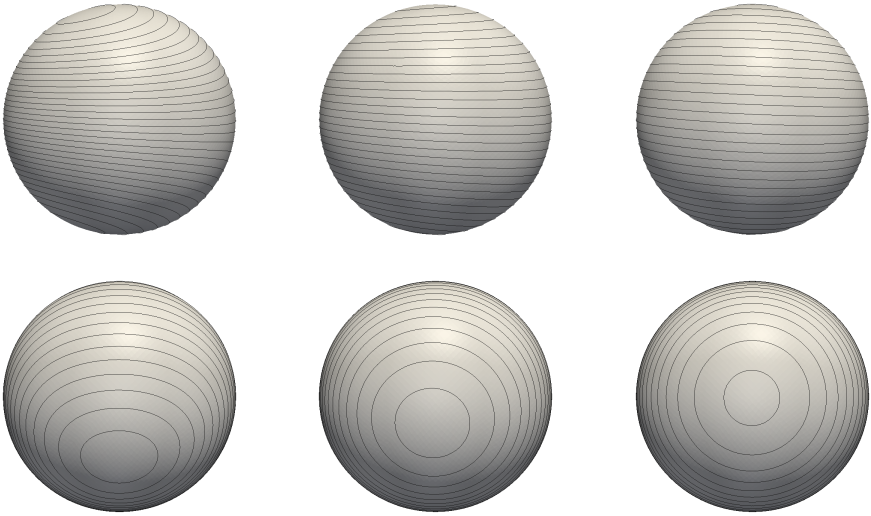

FIG. 1. *Evolution of streamlines for $t = 0, 0.25$ and $1$ (from left to right) for $\mu = 1.0$ on a sphere with $R = 1$ (top row: front view ; bottom row: top view).*

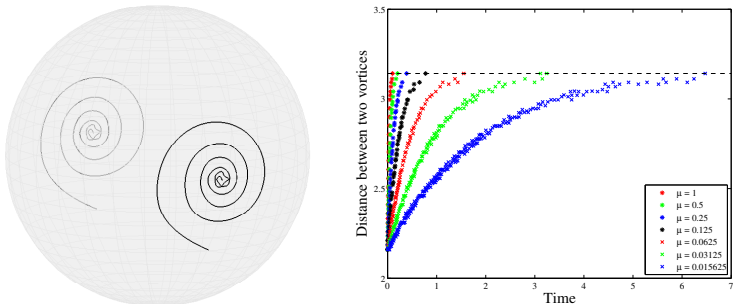

FIG. 2. *Vortex trajectories on the sphere for $\mu = 0.01$ (left) and distance between the two vortices over time for various viscosities (right).*

This stationary state is reached for all considered initial conditions and is clearly a result of the influence of the Gaussian curvature on the flow. However, a quantitative analysis of this influence is not possible for this geometry. To concentrate on, and better understand the geometric interaction, we therefore consider three simpler examples, which are adapted from [23]. The first considers a circular domain with a bump, slightly placed outside the center, the second a circular domain with a Gaussian saddle and the third the "Enneper disk", a minimal surface, which is characterized by a vanishing mean curvature $H$. Within the first two cases

$$\Gamma = \{(x_1, x_2, x_3)^T \in \mathbb{R}^3 \ : \ x_1^2 + x_2^2 < r^2, x_3 = h(x_1, x_2)\}$$

with a height-function $h$ specifying the bump

$$h(x_1, x_2) = \alpha r_0 \exp^{-\frac{(x_1-m_1)^2+(x_2-m_2)^2}{2r_0^2}}$$

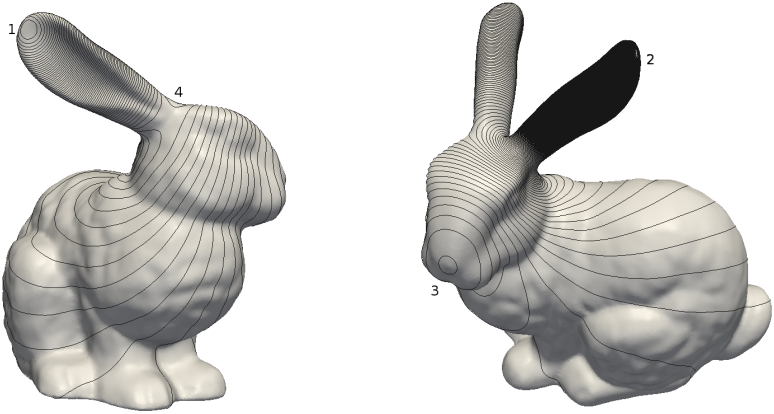

FIG. 3. *Streamlines on the Stanford bunny with numbering of the different defects. Numbers "1", "2" and "3" indicate vortices (+1 defects) and number "4" indicates a saddle (−1 defect).*

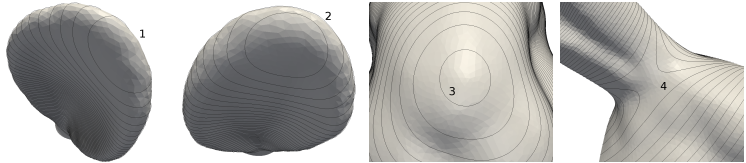

FIG. 4. *Identified defects with rescaled streamlines according to the marked positions in Fig. 3.*

with $\alpha = 2.5$, $r_0 = 0.2$ and position $(m_1, m_2)$, or the Gaussian saddle

$$h(x_1, x_2) = \frac{\alpha}{r_0}((x_1 - m_1)^2 - \lambda(x_2 - m_2)^2) \exp^{-\frac{(x_1-m_1)^2+(x_2-m_2)^2}{2r_0^2}}$$

with $\alpha = 1.5$, $\lambda = 0.99$, $r_0 = 0.2$ and position $(m_1, m_2)$, see Fig. 5. The "Enneper disk" is parameterized over the circular domain $\Gamma$ with $r = 1.5$ by

$$x = \frac{1}{3}(\frac{1}{3}x_1^3 - x_1 x_2^2 - x_1), \quad y = \frac{1}{3}(-\frac{1}{3}x_2^3 + x_2 x_1^2 + x_2), \quad z = \frac{1}{3}(x_1^2 - x_2^2),$$

see Fig. 5.

In all cases, we consider one +1 defect, which however does not result as a topological constraint, but is a consequence of the boundary condition, which we specify as

$$\phi = 2cr, \qquad \psi = c \quad \text{at} \quad \partial\Gamma \tag{3.8}$$

with a constant $c$. This induces a constant tangential velocity at the boundary and thus a vortex within $\Gamma$. We further specify zero initial conditions for $\phi$ and $\psi$.

Fig. 6 shows the stationary solutions. All results show the same qualitative influence by the Gaussian curvature as described for superfluids in [23].

In the first case, the bump leads to a lower velocity above the vortex (visible by a larger spacing between the contour lines), which creates a higher pressure and pushes the vortex away from the bump. Competing with the boundary condition, which favor the vortex to be at the center, this leads to a stationary profile with the vortex placed off the center. The second case is a vortex-trapping surface. The geometric potential has its absolute minimum at the center of the saddle, which attracts the vortex, independent of the position of the Gaussian saddle on the disk. The third case considers a minimal surface. In this example, the vortex is attracted to the middle of the surface. With this demonstration of the vortex-geometry interaction in an incompressible surface Navier-Stokes equation, we will now consider evolving surfaces with the goal to manipulate the flow field by changing the geometry.

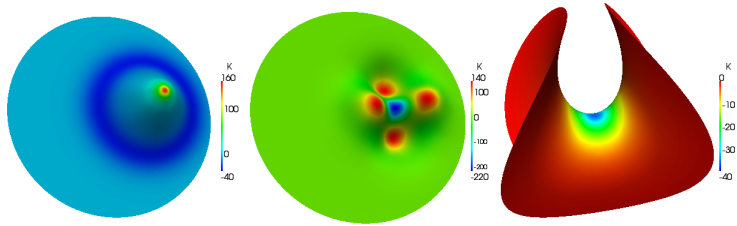

FIG. 5. *Considered geometries: (a) circular domain with a bump, (b) circular domain with a Gaussian saddle and (c) the "Enneper disk". The color coding is according to the Gaussian curvature $K$.*

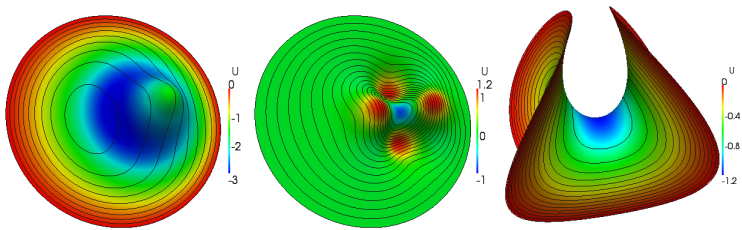

FIG. 6. *Stationary solution for: (a) circular domain with a bump, (b) circular domain with a Gaussian saddle and (c) the "Enneper disk". Shown are the contour lines for $\psi$. The color coding is according to the geometric potential $U_G$.*

**4. Evolving surface.** We consider a surface evolution in normal direction and assume that the surface area remains (at least globally) conserved. This requirement follows from our incompressibility assumption and is a typical constraint e.g. for lipid bilayer membranes [11]. We here specify the normal velocity, which is given by $\mathbf{v}_n = (0, 0, v_n)$ in the considered coordinate system. We thus obtain for the fluid velocity $\mathbf{u} = \mathbf{v}_3 + \mathbf{v}_n$ which has to satisfy $\nabla_\Gamma \cdot \mathbf{u} = \nabla_\Gamma \cdot \mathbf{v}_3 + \nabla_\Gamma \cdot \mathbf{v}_n = \nabla_\Gamma \cdot \mathbf{v} + v_n \nabla_\Gamma \cdot \mathbf{n} = \nabla_\Gamma \cdot \mathbf{v} + v_n H = 0$ with mean curvature $H$. Besides this modification of the incompressibility condition, the normal velocity also enters in the tangential balance of linear momentum, see [20] and [1, 19] for a formulation in the same notation as used here. We obtain

$$\partial_t \mathbf{v} + v_n H \mathbf{v} + \mathbf{v} \cdot \nabla_\Gamma \mathbf{v} = -\nabla_\Gamma p + \mu(\Delta_\Gamma^R \mathbf{v} + 2K\mathbf{v} - \nabla_\Gamma(v_n H))$$
$$- 2\mu \nabla_\Gamma \cdot (v_n \mathbf{S}) \quad (4.1)$$
$$\nabla_\Gamma \cdot \mathbf{v} + v_n H = 0 \quad (4.2)$$

on the evolving surface $\Gamma(t)$ with the shape operator $\mathbf{S} = \nabla_\Gamma \mathbf{n}$. The term including the mean curvature $H$ on the left hand side of eq. (4.1) follows from conservation of linear momentum and the transport theorem, see e.g. [7], whereas the term including the shape operator $\mathbf{S}$ is a consequence of the rate-of-deformation tensor $\mathbf{D} = \frac{1}{2}(\nabla_\Gamma \mathbf{v} + (\nabla_\Gamma \mathbf{v})^T) - v_n \mathbf{S}$, which includes the normal velocity and the shape operator and induces viscous tractions for non-uniform shape changes.

We rewrite the system as before as a surface vorticity-stream function formulation[1]

$$\partial_t \phi + \nabla_\Gamma \cdot (v_n H \nabla_\Gamma \psi) + J(\psi, \phi) = \mu(\Delta_\Gamma \phi + 2\nabla_\Gamma \cdot (K \nabla_\Gamma \psi)) \\ - 2\mu \nabla_\Gamma \cdot (\mathbf{n} \times (\nabla_\Gamma \cdot (v_n \mathbf{S}))) \quad (4.3)$$

$$\phi = \Delta_\Gamma \psi \quad (4.4)$$

and introduce the surface finite element approximation. Let $\Pi_h^m$ be a surface triangulation of $\Gamma(t^m)$ of mesh size $h$ and $\Gamma_h^m$ an interpolation of $\Gamma(t^m)$ such that

$$\Gamma_h^m = \bigcup_{Z \in \Pi_h^m} Z.$$

With the surface finite element spaces

$$V_h^m = \{v_h^m \in H^1(\Gamma_h^m) \mid v_{h|Z}^m \in P^1(Z), \forall Z \in \Pi_h^m\}$$

the surface finite element approximation now reads: Find $(\phi^m, \psi^m) \in V_h^m \times V_h^m$ such that for all $(\alpha, \beta) \in V_h^m \times V_h^m$

$$(d_\tau \phi^m, \alpha) - (v_n H \nabla_\Gamma \psi^m, \nabla_\Gamma \alpha) + (J(\psi^{m-1}, \phi^m), \alpha) = \\ \mu(-\nabla_\Gamma \phi^m - 2K\nabla_\Gamma \psi^m, \nabla_\Gamma \alpha) \\ -2\mu(\nabla_\Gamma \cdot (\mathbf{n} \times \mathbf{S}\nabla_\Gamma v_n), \alpha) \\ -2\mu(\nabla_\Gamma \cdot (\mathbf{n} \times (v_n \nabla_\Gamma \cdot \mathbf{S})), \alpha) \\ (\phi^m, \beta) = -(\nabla_\Gamma \psi^m, \nabla_\Gamma \beta)$$

where $K = K^m$, $H = H^m$, $\mathbf{S} = \mathbf{S}^m$ and $v_n = v_n^m$ the Gaussian and mean curvature, the shape operator of $\Gamma_h^m$ and the normal velocity at $t^m$, respectively. Again, these geometrical data are assumed to be given analytically or computable at the required accuracy, see Appendix A.

To demonstrate the approach, we modify the considered examples for the stationary circular domain and let the bump and the Gaussian saddle evolve. We first use $\alpha = \alpha(t)$ in the considered height profile with $\alpha(0) = 0$ and $\alpha(T) = 2.5$ or $1.5$, for the bump and Gaussian saddle, respectively. Fig. 7 shows the evolution of the streamlines, which adapt to the changing geometry leading to the same stationary solution as before. As a second example, we let the bump and the Gaussian saddle rotate around the center, with $(m_1, m_2) = (m_1(t), m_2(t))$, see Fig. 8 and Fig. 9.

**5. Conclusions.** The mathematical formulation of an incompressible fluid on a curved evolving surface is considered. The incompressible surface Navier-Stokes equation contains additional geometric terms which induce a strong coupling between topology, geometry and fluid properties. On closed surfaces, topological constraints might require the presence of defects in the flow field. These defects respond to the geometry of the surface and its changes and interact with each other. This leads to a highly nonlinear coupling, which can induce non-uniform surface flow as a response of lateral motion and thus opens new possibilities to manipulate surface flow. We demonstrate this interplay on simple examples using a surface vorticity-stream function formulation, which is solved by using parametric finite elements. Within the numerical treatment we assume all geometric quantities, such as the mean and Gaussian curvature, $H$ and $K$, the shape operator $\mathbf{S}$ and the normal velocity $v_n$, to be given analytically or computable at the required accuracy. For the example of the Stanford bunny we are using gradient recovery strategies to approximate the shape operator and the mean and Gaussian curvature. For more general surfaces or generalizations of the model in which $v_n$ is not specified but follows from the balance of linear momentum normal to the surface, the computation of $H$, $K$, $\mathbf{S}$ and $v_n$ requires more care, see Appendix A.

---

[1] We here used the identity $\text{rot}(w) = -\nabla_\Gamma \cdot (\mathbf{n} \times w)$ with a vector field $w$ defined on $\Gamma(t)$ for the additional term on the right hand side of eq. (4.1). All other terms are treated as proposed before or in [17].

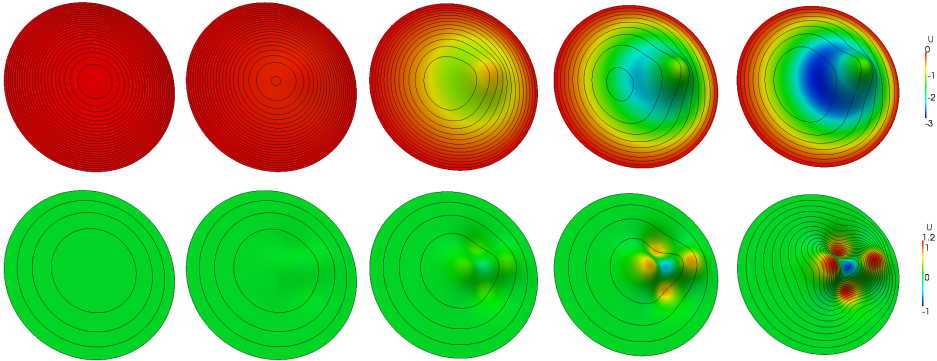

FIG. 7. *Time evolution of the bump and Gaussian saddle for $t = 2, 14, 26, 38$ and $50$. Shown are the contour lines for $\psi$. The color coding is according to the geometric potential $U_G$.*

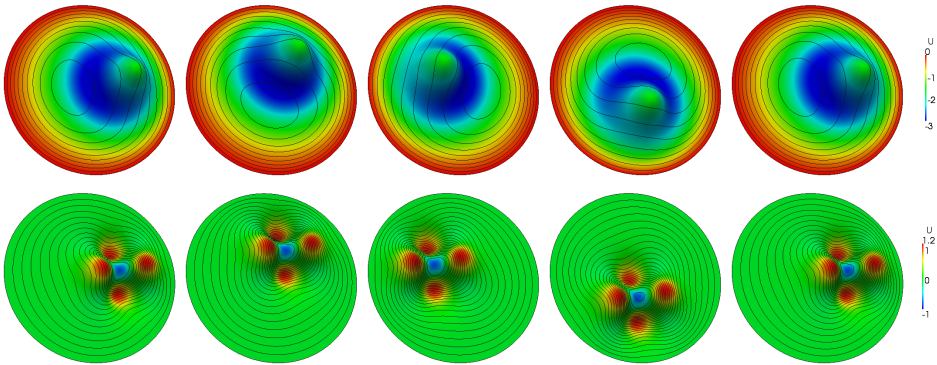

FIG. 8. *Time evolution of the bump and Gaussian saddle for $t = 2, 11, 19, 27$ and $36$. Shown are the contour lines for $\psi$. The color coding is according to the geometric potential $U_G$.*

## Appendix A. Approximation of geometric quantities on surfaces.

The numerical solution of the considered surface vorticity-stream function formulation requires various geometric quantities of the surface $\Gamma(t)$. If $H$, $K$ and $\mathbf{S}$ of the surface $\Gamma(t)$ cannot be computed analytically, a point-wise approximation of these quantities is needed and has to be computed from the discretized surfaces $\Gamma_h^m$. Vari-ous numerical approaches have been proposed for this task. We here refer to [10] for a parametric finite element approach, for which convergence of order $k$ could be shown in the $L_2$-norm for Lagrange elements of polynomial degree $k + 1$. The approach is based on a discrete definition of $\mathbf{H} = -\Delta_\Gamma \mathrm{id}$ and $\mathbf{S} = \nabla_\Gamma \mathbf{n}$, with curvature vector $\mathbf{H} = H\mathbf{n}$ and the formula for integration by parts on $\Gamma$

$$\int_\Gamma \nabla_\Gamma f \, d\Gamma = -\int_\Gamma f\mathbf{H} \, d\Gamma.$$

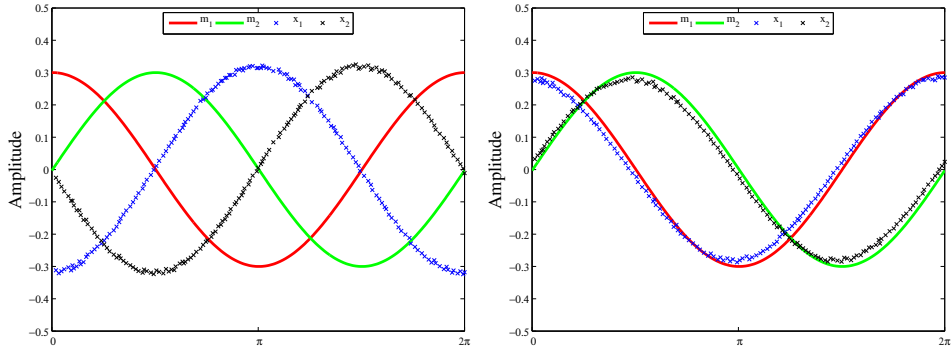

FIG. 9. *Evolution of the bump location (left) and the Gaussian Saddle location (right) denoted by $(m_1, m_2)$ and the appropriate vortex location $(x_1, x_2)$ for a full rotation period.*

The surface finite element approximation reads: Find $(\mathbf{H}_h, \mathbf{S}_h) \in (V_h^m)^3 \times (V_h^m)^{3\times 3}$ such that for all $(\gamma, \delta_1, \delta_2, \delta_3) \in (V_h^m)^3 \times (V_h^m)^3 \times (V_h^m)^3 \times (V_h^m)^3$

$$(\mathbf{H}_h, \gamma) = (\nabla_\Gamma \mathrm{id}, \nabla_\Gamma \gamma) \tag{A.1}$$

$$(\mathbf{S}_{1,h}, \delta_1) = -(n_1, \nabla_\Gamma \cdot \delta_1) - (n_1, \mathbf{H}_h \cdot \delta_1) \tag{A.2}$$

$$(\mathbf{S}_{2,h}, \delta_2) = -(n_2, \nabla_\Gamma \cdot \delta_2) - (n_2, \mathbf{H}_h \cdot \delta_2) \tag{A.3}$$

$$(\mathbf{S}_{3,h}, \delta_3) = -(n_3, \nabla_\Gamma \cdot \delta_3) - (n_3, \mathbf{H}_h \cdot \delta_3) \tag{A.4}$$

with $\mathbf{S}_h = (\mathbf{S}_{1,h}, \mathbf{S}_{2,h}, \mathbf{S}_{3,h})^T$ and $\mathbf{n} = (n_1, n_2, n_3)^T$ in the Cartesian coordinate system $(\mathbf{e}_1, \mathbf{e}_2, \mathbf{e}_3)$ and the finite element spaces $V_h^m = \{v_h^m \in H^1(\Gamma_h^m) \mid v_{h|Z}^m \in P^{k+1}(Z), \forall Z \in \Pi_h^m\}$. The approximation of the mean and Gaussian curvature $H_h$ and $K_h$ can now be computed from the eigenvalues of $\mathbf{S}_h$.

The present implementation uses Lagrange elements of polynomial degree 1, but local gradient recovery strategies at each quadrature point for the example of the Stanford bunny, which qualitatively leads to the same results as for elements of higher degree in [10]. A detailed convergence study and comparison with other approaches will be discussed elsewhere [16].

**Acknowledgement.** This work was supported by DFG within SPP 1506 through Vo899/11-2. We would like to thank Simon Praetorius for fruitful discussions and his support.